\documentclass{elsart}

\usepackage{natbib}
\usepackage{epsfig}
\usepackage{amssymb}

\begin{document}

\begin{frontmatter}

\title{X-ray Variability of Active Galactic Nuclei and the Flare Model}

\author[France]{R.W. Goosmann},
\ead{rene.goosmann@obspm.fr} 
\author[Poland]{B. Czerny},
\author[France]{A.-M. Dumont},
\author[France]{M. Mouchet},
\author[Poland]{A. R\'o\.za\'nska},
\author[Chech1,Chech2]{V. Karas}, \and
\author[Chech2]{M. Dov{\v c}iak}
\address[France]{LUTH, Observatoire de Paris, Meudon, 5 place Jules
  Janssen,\\ 92190 Meudon, France}
\address[Poland]{Nicolaus Copernicus Astronomical Center, Bartycka
  18,\\ 00-716 Warsaw, Poland}
\address[Chech1]{Charles University, Faculty of Mathematics and
  Physics, Ke Karlovu 3,\\ 121 16 Prague 2, Czech Republic}
\address[Chech2]{Astronomical Institute, Academy of Sciences, Narodni
  3,\\ 117 20 Prague 1, Czech Republic}

\begin{abstract}
Short-term variability of X-ray spectra has been reported
for several Active Galactic Nuclei. Significant X-ray flux variations
are observed within time scales down to $10^4$\textemdash$10^5$
seconds. We discuss short variability time scales in the framework of
the X-ray flare model, which assumes the release of a large hard X-ray
flux above a small portion of the accretion disk. The resulting
observed X-ray spectrum is composed of the primary radiation as well
as a reprocessed/Compton reflection component that we model performing
numerical radiative transfer simulations. We conduct Monte-Carlo
simulations of large flare distributions across the disk including
relativistic corrections of the observed radiation and discuss general
implications on AGN spectra and their variability.
\end{abstract}


\end{frontmatter}

\section{Introduction}
\label{intro}

The X-ray emission of Active Galactic Nuclei (AGN) can be explained by two components: a primary radiation emerging from a hot, optically thin medium and a reprocessed component coming from colder, optically thick medium. The existence of these two media can be explained by the flare model, which was first suggested by \citet{galeev1979} and developed in many subsequent papers for both galactic black holes and AGN. The AGN flares are assumed to be driven similarly to solar flares: magnetic reconnection creates a compact source above the accretion disk, the source emits the primary spectrum partly reaching the observer directly and partly illuminating the disk atmosphere, the atmosphere then emits the reprocessed spectrum.

\begin{figure}[h]
  \begin{center}
    \epsfxsize=6cm
    \epsfbox{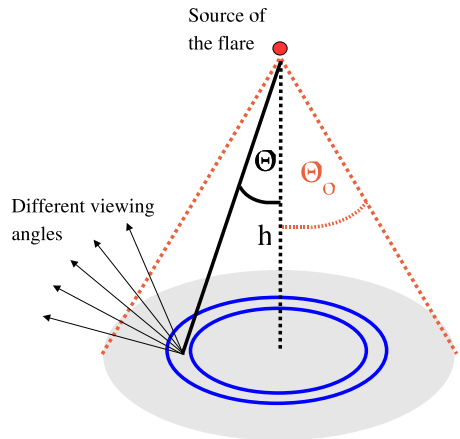}
  \end{center}
  \caption{Flare geometry}
  \label{flaregeometry}
\end{figure}

In this paper we present the first modeling results for the reprocessed radiation coming from the irradiated area, the so-called hot-spot, underneath a compact flare source as indicated in Fig.~\ref{flaregeometry}. We take into account details of the flare geometry including the dependence on the incident angle of the primary radiation.

We further conduct Monte-Carlo simulations of random flare
distributions across the accretion disk in order to construct rms
X-ray spectra. These simu\-lations also contain relativistic
corrections due to the curved space-time in the vicinity of the black
hole. We show general trends for several parameters such as the radial
distribution of hot-spots and of the incident radiation flux over the
disk.

\section{Modeling}
\label{modelspot}

\subsection{The reprocessed spectra across the hot-spot}

We perform detailed radiative transfer using the codes \emph{TITAN}
and \emph{NOAR} described in \citet{dumont2000} and in
\citet{dumont2003}. \emph{TITAN} conducts radiative transfer inside
optically thick media, determining the temperature and ionization
structure. \emph{NOAR} is a Monte-Carlo code treating the Compton
processes. We consider a plane-parallel section of the accretion disk
atmosphere at $18\:R_g$ ($R_g =  \frac{GM}{c^2}$, $M$ being the mass of
the black hole) from the disk center. The hydrostatic equilibrium of
such an atmosphere is calculated following \citet{rozanska1999}
assuming a $10^8$ solar mass black hole and an Eddington accretion
rate of 0.001. \Citet{collin2003} have shown that a change of the
hydrostatic equilibrium due to the onset of the flare happens slowly
compared to the adjustment of the temperature and ionization profile
inside the medium. We therefore assume that the density structure of
the medium does not change for a significant fraction of the total
flare duration.

For the incident flux $F_{inc}$ we presume a power law with $F_{inc} \propto E^{-\alpha}$, $\alpha = 0.9$, between 0.1 eV and 100 keV. The ratio $F_{inc}/F_{disk}$ of the incident flux to the flux coming from the disk is fixed at  $144$, hence the flare by far dominates over the disk emission.

It can be seen from Fig.~\ref{flaregeometry} that the incident angle and the ionization parameter $\xi=\frac{4 \pi F_{inc}}{n_H}$ of the primary radiation (with $F_{inc}$ being the incident flux and $n_H$ being the hydrogen density at the disk surface) depend on the position within the hot-spot. This aspect has not yet been considered in previous flare models where the disk illumination was defined as semi-isotropic \citep{czerny2004a,collin2003} or with a constant incident angle \citep{nayakshin2001}.

For reasons of energy equipartition the distance $h$ of the flare
source above the disk surface should be similar to the pressure scale
height of the disk \citep{galeev1979}. Applying equation (2.8) of
\citet{shakura1973} to our case we derive $h \sim 1.9 \times 10^{11}$
cm $\sim 0.013\:R_g$. We define a half-opening angle $\Theta_0=60^\circ$
for the flare cone, which, for an isotropically emitting source,
corresponds to $50\%$ of the incident flux reaching the disk being
reprocessed. Then, the  hot-spot radius of $h \times \tan 60^\circ
\sim 0.023\:R_g$ verifies our implicit assumption that the hydrostatic
equilibrium computed at $18\:R_g$ is sufficiently accurate across the
whole hot-spot.

The observed spectrum is constructed as the integrated spectra of
successive concentric rings at incident angles with
$0^\circ<\Theta<\Theta_0$. The spectra can be evaluated at various
disk viewing angles $i$ (measured from the disk symmetry
axis). Details of the calculation method will be published in a
following article \citetext{Goosmann et al., in preparation}.

\subsection{The rms-variability of random flare distributions}

To investigate the consequences of this more detailed modeling of the hot-spot we perform Monte-Carlo simulations of many flares distributed over the accretion disk. The procedure is outlined in \citet{czerny2004a}. We assume the hot-spot emission to be independent of the flare position. Radial probability distributions $p(R_i,\phi_i)$ for the appearance of a flare can be defined as

\begin{equation}
   p(R_i,\phi_i)
      \equiv p(R_i) p(\phi_i)
      = \frac{(\gamma_{rad}+2)R_i^{\gamma_{rad+1}}}
             {R_{out}^{\gamma_{rad+2}}-R_{in}^{\gamma_{rad+2}}}
        \times \frac{1}{2\pi}.
   \label{raddist}
\end{equation}

Herein $R_i$ and $\phi_i$ are polar coordinates, $R_{in}$ and
$R_{out}$ denote the inner and the outer disk radius. For
$\gamma_{rad} = 0$ the hot-spot distribution is uniform while for
positive (negative) values of $\gamma$ the number of flares rises in
the outer (inner) disk part. We also enable a radial dependence of the
hot-spot incident flux:

\begin{equation}
   F_{inc} = F_0 \left[ \frac{R_i}{R_{in}} \right]^{-\beta_{rad}}.
   \label{fluxstrength}
\end{equation}

Since the dissipation within a Keplerian disk scales with $r^{-3}$ we expect $\beta_{rad}$ of the order of 3. Finally, the radial dependence of the flare life-time is chosen to scale with the Keplerian time scale.

The average number of hot-spots existing on the disk is denoted by $n_{mean}$, the mean X-ray luminosity of the primary source by $L_X = 8 \times 10^{43}$~erg~s$^{-1}$. The rms-spectra are computed from $n_{seq} = 50$ consecutive model observations, each of an integration time $T_{integ} = 6156$~s. Relativistic effects are included by application of the code \emph{KY} \citep[see][for a code description]{dovciak2004}. For the following computations we assume a Kerr black hole of $10^7$ solar masses with spin parameter $a = 0.95$. The inner disk radius is at $1.8\:R_g$, the outer one at $50\:R_g$. The viewing angle of a distant observer is taken to be $i = 30^\circ$.

\section{Results}
\label{results}

\subsection{The reprocessed spectra across the hot-spot}

\begin{figure}[h]
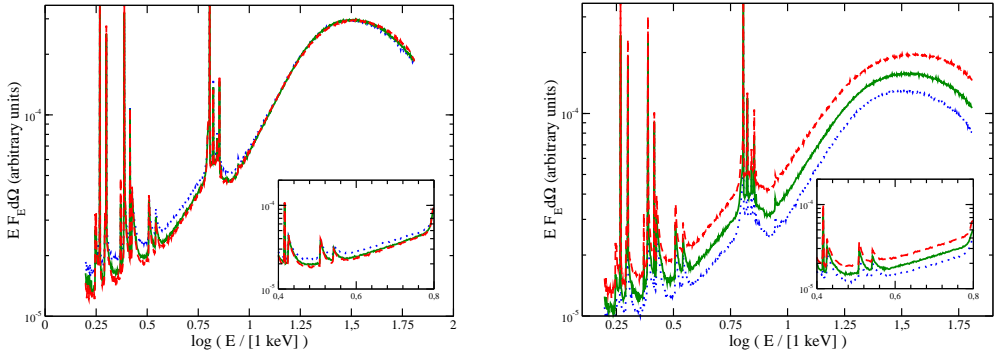

   \begin{center}
      \vskip 0.5cm
      \epsfxsize=6cm
      \epsfbox{ring-spectV13.eps}
      \hskip +1cm
      \epsfxsize=6cm
      \epsfbox{ring-spectV62.eps}
   \end{center}
   \caption{Reprocessed spectra for different incident angles of the
            irradiating hard X-rays. Dotted line: $\cos{\Theta} = 0.95$
            $(\Theta=18^\circ)$. Solid line: $\cos{\Theta} = 0.65$
            $(\Theta=49^\circ)$. Dashed line: $\cos{\Theta} =
            0.35$ $(\Theta=70^\circ)$. Left: face-on view. Right:
            intermediate viewing angle at $62^\circ$ from the
            normal. The inlets show a zoom between $0.4$ and $0.8$ keV.}
   \label{ringspect}
\end{figure}

In this paper we only examine the pure reprocessed component
(e.g. without dilution by the primary radiation) for various positions
within the hot-spot, e.g. for various incident angles of the
irradiating flux. In Fig.~\ref{ringspect} we show the reprocessed flux
$EF_E$ versus photon energy $E$ at several incident angles
$\Theta$. The results are shown for a face-on view, as assumed in
Seyfert-1 galaxies, and for an intermediate viewing angle, as supposed
for Seyfert-2 galaxies. For both viewing angles spectral variations
are found in the soft X-rays across the hot-spot. These changes are
more significant at higher viewing angles and affect the whole
spectrum. Note that the dependence on the incident angle of the
primary spectrum is opposite for the two viewing angles shown. At
face-on view the flux decreases with rising incident angle, while at
intermediate viewing angles it rises.

\begin{figure}[h]
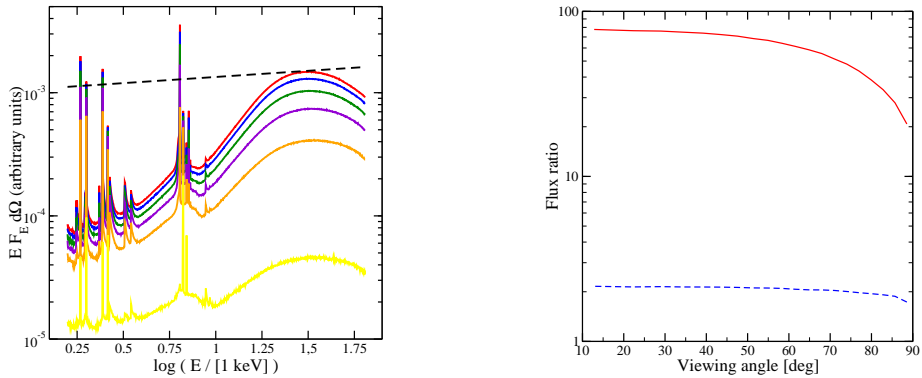

   \begin{center}
      \vskip 0.5cm
      \epsfxsize=5cm
      \epsfbox{flarespectraHA60.eps}
      \hskip +2cm
      \epsfxsize=5cm
      \epsfbox{fluxratios.eps}
   \end{center}
   \caption{Left: integrated spectra of a hot-spot with half-opening
            angle of $60^\circ$. The viewing angles increases from top
            ($i = 13^\circ$) to bottom ($i = 89 ^\circ$). Right: flux
            ratios versus viewing angle for such a hot-spot. Dashed
            line: F[5.5keV]/F[4keV] (soft X-ray slope). Solid line:
            F[28keV]/F[4keV] (hard/soft X-ray slope).}
   \label{integspect-ha60}
\end{figure}

The spectra emitted by the whole hot-spot are shown in
Fig.~\ref{integspect-ha60}. For increasing viewing angles the spectra
decrease systematically, as does the soft X-ray slope and the ratio of
the Compton hump to the soft X-ray spectrum.

\subsection{The rms-variability of random flare distributions}

\begin{figure}[h]
   \begin{center}
      \epsfxsize=6.5cm
      \epsfbox{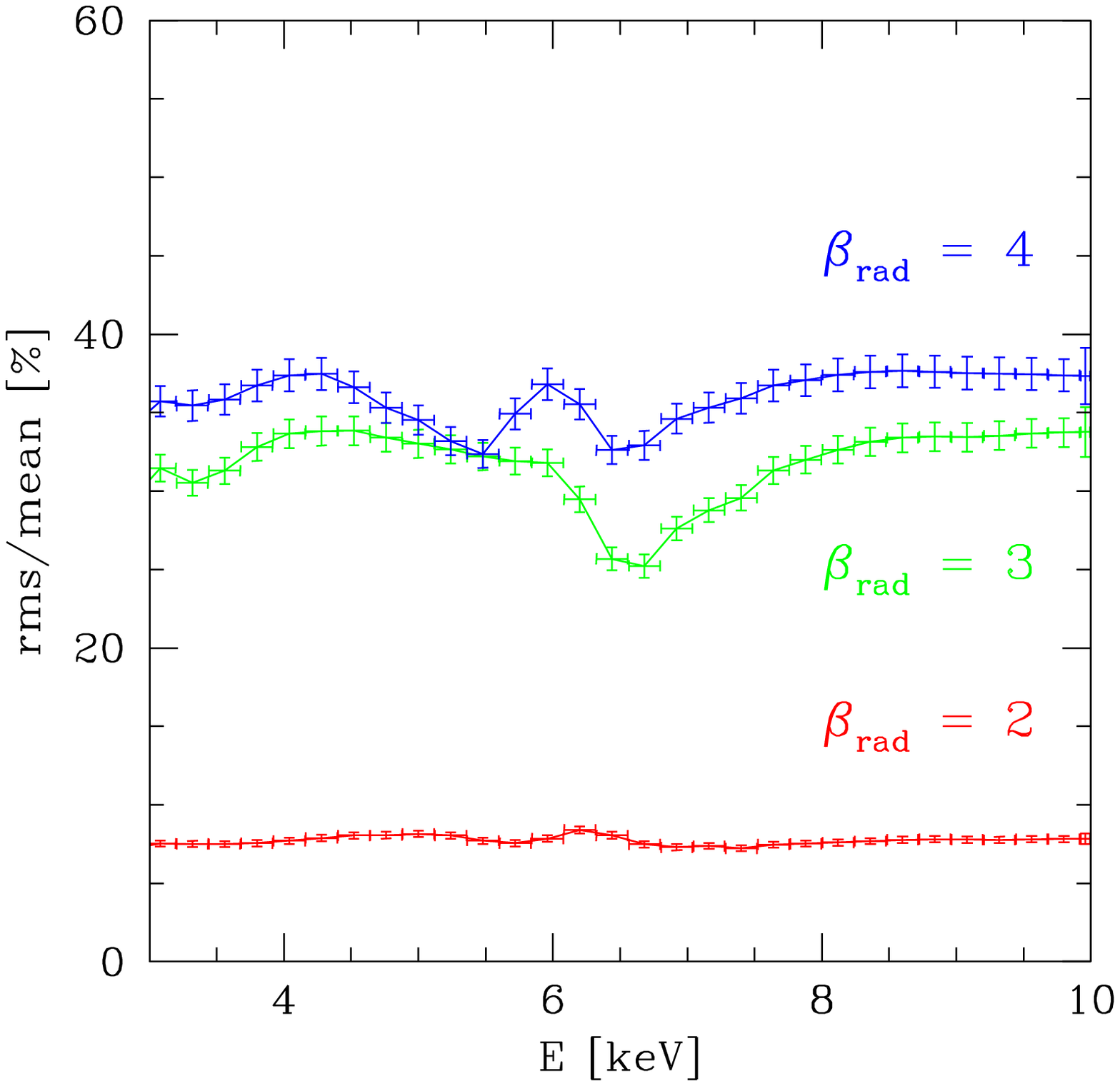}
      \epsfxsize=6.5cm
      \epsfbox{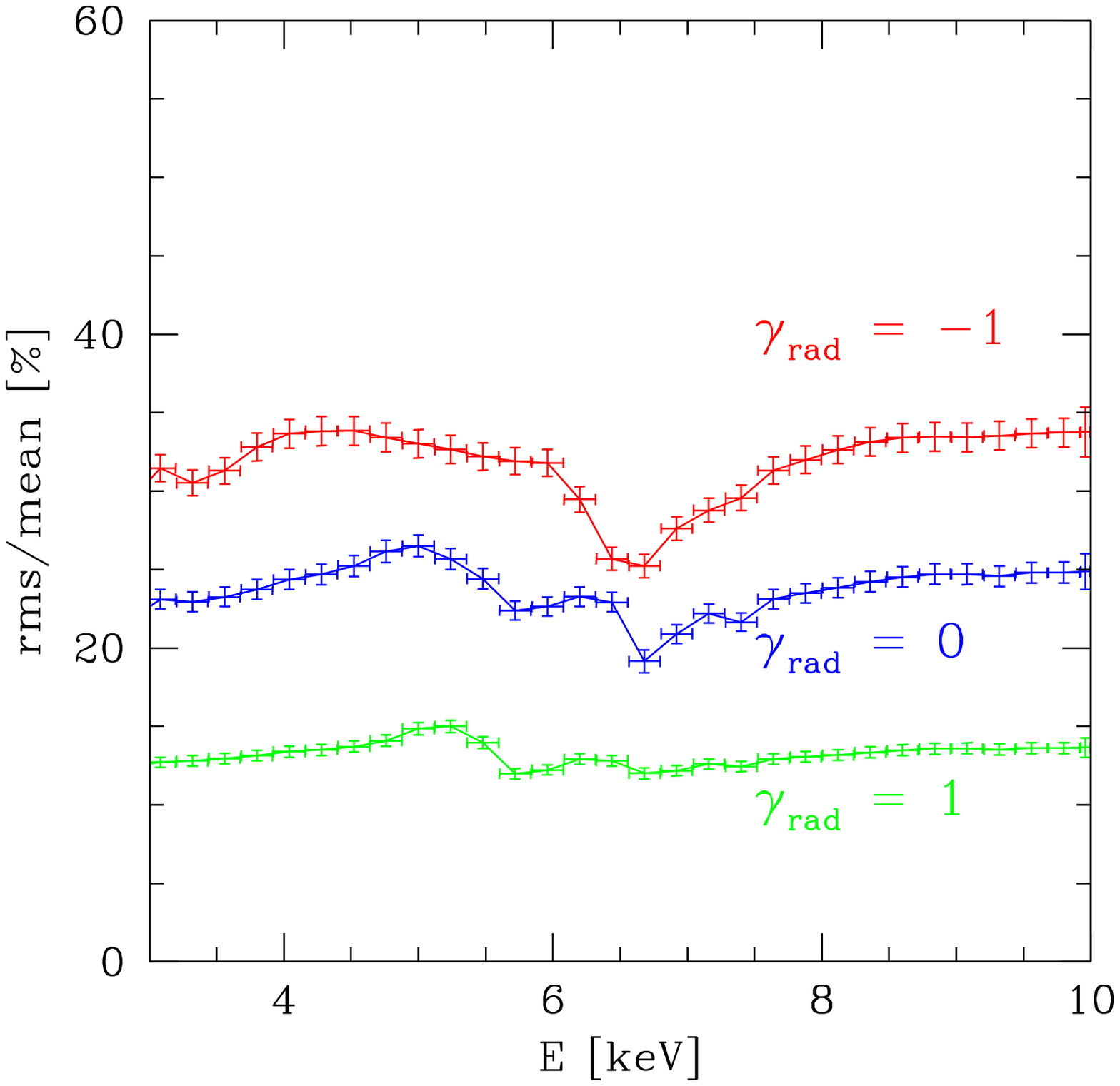}
   \end{center}
   \caption{Left: Dependence of the rms/mean spectrum on $\beta_{rad}$
   (other model parameters: $\gamma_{rad} =  -1.5$, $n_{mean} =
   300$). Right: Dependence of the rms/mean spectrum on $\gamma_{rad}$
   (other model parameters: $\beta_{rad} =  3$, $n_{mean} = 100$).}
   \label{rms}
\end{figure}

In Fig.~\ref{rms} (left) we show the dependence of the rms spectra on
the radial distribution of the incident radiation flux. Again, we do
not consider any dilution by the primary radiation of the flare. We
assume that the hot-spots at outer radii are dimmer, but if the
decrease of the luminosity is fast, the rms spectrum has a much more
complicated structure since it depends strongly on a few innermost
brightest hot-spots. Computing various random realizations of the
process without any variations in the model parameters, we see
statistical differences between them. If the hot-spot's surface
brightness decreases more slowly the role of the outer hot-spots
increases, causing a major drop in the overall variability level.

In Fig.~\ref{rms} (right) the rms spectra for several radial hot-spot distributions are shown. The value of $\gamma_{rad}=0$ corresponds to a random but uniform coverage of the disk surface by the hot-spots. If the flares and hot-spots appear more frequently in the outer part of the disk ($\gamma_{rad}> 0)$, the overall variability decreases but a variable red wing is still present. If the flares appear relatively more frequently in the inner part of the disk ($\gamma_{rad})< 0$, a clear dip is seen at the position of the unshifted iron line.

\section{Discussion and Conclusions}
\label{dicussconclude}

Modeling the reprocessed radiation of the hot-spot reveals that the pure reflection component varies with the position in the hot-spot. This variation is stronger for intermediate (edge-on) viewing angles than for a face-on viewing direction. The effect is caused by the $\xi$ parameter varying with the distance from the hot-spot center. For outer parts of the hot-spot $\xi$ is lower, it will thus cause a colder medium and a thinner hot surface of the atmosphere. The temperature of the medium determines its absorption properties, while the hot surface will influence, by electron scattering, the angular dependence of the re-emitted flux. The incident angle of the radiation also has an impact on the spectral shape as the Compton scattering phase function is not isotropic. 

The integrated spectra show a strong dependence on the viewing direction. This effect is partly explained by the size of the projected surface area of the hot-spot seen at different inclinations. However, there also is a dependence of the overall spectral slope on the viewing direction that could be used to put observational constraints on the disk inclination.

The modeling of rms spectra for distributions of many flares across the disk reveals that flares at the inner part of the disk dominate the variability. Either a strong concentration of flares at the inner disk part or a relatively higher luminosity of inner disk flares increases the level of variability. Due to the light bending, an orbiting flare in the central region of the disk is effectively seen at various local viewing angles, which increases spectral variations. Although the overall level of variability is only weakly dependent on the energy, there are variations of the rms spectra for the energy range of the K$\alpha$ line. This effect is due to the strong gravity close to the Kerr black hole. It could give a hint to the lack of variability in the iron K$\alpha$ line observed in MCG -6-30-15 \citep[see e.g.][and references therein]{fabian2003}.

A basic assumption of our modeling is that the density structure of
the medium stays constant after the onset of the flare. Hence, we give
results for flares having a short duration compared to the dynamical
time scale. On longer time scales the medium expands to obtain a new
hydrostatic equilibrium \citep[see][for details]{czerny2004b} and the
spectral properties change.

\end{document}